\documentclass{aastex}

\usepackage{spr-astr-addons}

\usepackage{graphicx}

\begin{document}

\title{The Big Trip and Wheeler-DeWitt equation}
\shorttitle{The Big Trip and Wheeler-DeWitt equation}
\shortauthors{A.V. Yurov et al.}

\author{Artyom V. Yurov} \and \author{Artyom V. Astashenok}
\affil{Baltic Federal University of I. Kant, Theoretical Physics Department,
 Al.Nevsky St. 14, Kaliningrad 236041, Russia}
\and
\author{Valerian A. Yurov}
\affil{University of Missouri, Department of Mathematics, 202 Mathematical Sciences Bldg,
 Columbia, MO, 65211 USA}
%\email{\emaila}

\date{\today}

\begin{abstract}

Of all the possible ways to describe the behavior of the universe that has undergone a big trip the Wheeler-DeWitt equation should be the most accurate -- provided, of course, that we employ the correct formulation. In this article we start by discussing the standard formulation introduced by Gonz\'alez-D\'iaz and Jimenez-Madrid, and show that it allows for a simple yet efficient method of the solution's generation, which is based on the Moutard transformation. Next, by shedding the unnecessary restrictions, imposed on aforementioned standard formulation we introduce a more general form of the Wheeler-DeWitt equation. One immediate prediction of this new formula is that for the universe the probability to emerge right after the big trip in a state with $w=w_0$ will be maximal if and only if $w_0=-1/3$.

\end{abstract}


\section{\label{sec:level1}Introduction}

In recent years the mathematical community of cosmologists has provided us with a multitude of new amazingly original models, not the least of which is the ``Phantom fields'' concept, that (among other things) notably postulates the violation of the weak energy condition (WEC) $\rho>0$, $\rho+p/c^2>0$ (\cite{Caldwell-1}, \cite{Caldwell-2}), where $\rho$ is the fluid density and $p$ is the pressure (for recent reviews, see \cite{Dark-1}, \cite{Dark-2}, \cite{Dark-4}, \cite{Dark-5}, \cite{Dark-6}). Such fields, as follows from quantum theory \cite{Carrol}, should manifest themselves via the scalar field with the negative kinetic term. The thorough investigation asserts that fields with such a property should be anything but fundamental. Still, one cannot guarantee that no Lagrangian with the negative kinetic terms would arise to serve as an {\em effective} model. And that, indeed, is exacly what happens in at least some of the supergravity models (\cite{Nilles}), as well as in the gravity theories with highest derivatives (\cite{Pollock}) and even in the field string theory (\cite{Aref'eva}) (in the models describing the fermion NSR-string with regard for GSO$_-$-sector (see also \cite{Sen})). Lately, the ``phantom energy'' had even managed to penetrate into the domain of yet another ``exotic'' cosmological model -- the brane theory (see, for example, \cite{Sahni}, \cite{Yurov-1}).
\newline
{\bf Remark 1}. Although most theoreticians believe that the phantom energy is nothing but an effective model, this shared notion as of yet remains to be backed up by anything rigorous. In fact, there are some evidence of quite the opposite. One particular example is the phenomena of ''crossing of the phantom division line'', first discovered and discussed in \cite{Yurov-2}, \cite{ENO}, \cite{Yurov-3}, and later interpreted in \cite{Andrianov}. This ``crossing'' is smooth and it is an inherent property of some of exact solutions of Einstein equations. One can even argue that the smooth (de)-phantomization is a sufficiently general property of Einstein equations. Therefore if we were (following \cite{Andrianov}) guided by a belief that Einstein equations for gravity are more fundamental than the concrete form of the Lagrangian for other fields, then one can conclude that the ``crossing of the phantom division line'' is a new physically plausible fundamental property of gravitation.

The particular interest to models with the phantom fields is caused by their prediction of so-called "Cosmic Doomsday" also known as
big rip \cite{Caldwell-1}, \cite{Starobinsky}, \cite{Frampton}, \cite{Nesseris}, \cite{Nojiri}, \cite{Gonzalez-1}. In case of the phantom
energy we have $w=p/(c^2\rho)=-1-\epsilon$ with $\epsilon>0$.
Integration of the Einstein-Friedmann equation for the flat universe results in
\begin{equation}
\begin{array}{cc}
\displaystyle{
a(t)=\frac{a_0}{\left(1-\xi t\right)^{2/3\epsilon}},}\\
\displaystyle{
\rho(t)=\rho_0\left(\frac{a(t)}{a_0}\right)^{3\epsilon}=\frac{\rho_0}{(1-\xi
t)^2}}, \label{1}
\end{array}
\end{equation}
where $\xi=\epsilon\sqrt{6\pi G\rho_0}$. We choose $t=0$ as the
present time, $a_0\sim 10^{28}$ cm and $\rho_0$ to be the present
values of the scale factor and the density. There, if
$t=t_*=1/\xi$, we automatically get the big rip.
\newline
{\bf Remark 2}. Even though we are certainly uncomfortable with the conclusion that the universe might be doomed to be (literally) blown up in the big rip, there is simply not too many options available to avert or delay it. So far, we only know of four:

(i) treating phantom energy as nothing more than an effective model (see above);

(ii) using the quantum effects to delay the singularity (\cite{Nojiri-2});

(iii) introducing a new time variable such that the big rip singularity will be delayed indefinitely (i.e. its occurrence will be pushed to $t\to\infty$) (\cite{Gonzalez-2});

(iv) the last and most intriguing of all: avoiding the big rip by means of another comparatively ''big'' cosmological effect -- the so-called big trip (see below).

In \cite{Gonzalez-3} Pedro F. Gonz\'alez-D\'iaz, and Jose A. Jime-nez-Madrid had introduced a hypothesis that a smooth exit from the phantom inflationary phase might actually be achievable provided one takes into account a multiversity scenario in which, right before reaching the big rip, the primordial phantom universe travels in time towards the future state, ultimately landing in a state where it will be filled with just usual radiation and no phantom fields. This transition had been called ``the big trip'' and it had been assumed it would take place due to a sort of anthropic principle, reserving the role of the final destination (of the time transition) for our current universe. It has been shown in \cite{9} that if $p =-(1+\epsilon)c^2\rho$ is a cosmological fluid's equation of state, then
\begin{equation}
c{\dot b}(t)=2\pi^2\epsilon GD\rho(t)b^2(t), \label{dotb}
\end{equation}
where $b(t)$ is the throat radius of a Morris-Thorne wormhole and $D$ is the dimensionless quantity. According to \cite{9} we can choose $D\sim 4$ (This is true only if $w<0$. If $0<w\le 1$ then $D\sim
A=(1+3w)^{(1+3w)/2w}/(4w^{3/2})$; see
 \cite{Babichev}). The equation (\ref{dotb}) describes the evolution of $b(t)$ with respect to the phantom
energy's accretion. Integration of the (\ref{dotb}) gives us
\begin{equation}
\frac{1}{b(t)}=\frac{1}{b_0}-\frac{2\pi^{2}\epsilon\rho_{0}GDt}{c(1-\xi t)}. \label{1b}
\end{equation}
Therefore at
\begin{equation}
{\tilde t}=\frac{c}{\epsilon(c\sqrt{6\pi G \rho_{0}}+2\pi^{2}\rho_{0}b_{0}G D)} \label{tild}
\end{equation}
we get $b({\tilde t})=\infty$. As we can see ${\tilde t}<t_*$, and therefore this universe will indeed turn achronal well before the occurrence of the big rip. In accord to \cite{9}, in the process of phantom energy's accretion, for the time period starting right after the radius of the wormhole $b(t)$ exceeds the scale factor $a(t)$ and stretching up to $\tilde t$, the wormhole becomes an active Einstein-Rosen bridge which can, in principle, be used to escape the big rip.
%%%%%%%%%%%%%%%%%%%%%%%%%%%%%%%%%%%%%%%%%%%%%%%%%%%%%%%%%%
\newline
{\bf Remark 3}. The notion of big trip has had its share of criticism - the article \cite{Faraoni} of Faraoni is by far the most recent and thorough one among those expressing the doubts about the model. However, the objections of Faraoni has been defended against in a response article \cite{Gonzalez-4}, where the detailed answers to all the objections were given. It is interesting to note that even phantom models with no big rip singularity do admit wormhole solutions
and hence allow for a possibility of big trip via such a wormhole (\cite{Astashenok}).

The description of big trip can be executed with help of Wheeler-DeWitt equation for wave function of the universe. To derive it authors of \cite{Gonzalez-3} had introduced the spacetime manifold M for a flat FRW universe with metric
\begin{equation}\label{PedroMetric}
ds^2=-N^2 dt^2 + a(t)^2 ~d\Omega_3^2,
\end{equation}
where $N$ is the lapse function arising due to foliation of the manifold $M$, $d\Omega_3^2$ is the metric on the unit threesphere, and $a(t)$ is the scale factor. Next, they had assumed that the parameter of equation of state $w$ is time-dependant, but in such a way that at any moment of time we should have
\begin{equation}\label{restrict}
\ddot{w}=0,
\end{equation}
so the scalar curvature $R$ would have to be generalized by
\begin{equation} \label{hatR1}
R\to {\hat R}=R-\frac{{\dot a}{\dot w}\ln a^9}{a},
\end{equation}
where ``dot'' indicates differentiation with respect to time. Now consider the universe filled with scalar field $\phi$ with the Lagrangian
$$
L=\frac{1}{2}\left(\partial_{\mu}\phi\right)^2-V(\phi)=p=w\rho.
$$
The action integral of the manifold $M$ with boundary $\partial M$ will have the form
\begin{equation}
S=\int_Md^4x\sqrt{-g}\left(-\frac{{\hat R}}{2}+w\rho\right)-\int_{\partial M}
d^3x\sqrt{-h}{\rm Tr}{\hat K}, \label{action}
\end{equation}
where $K$ is the extrinsic curvature, $g={\rm det}g_{\mu\nu}$ and $h$ is the determinant of the general threemetric on a given hypersurface at the boundary $\partial M$. Note, that herein we are using the natural system of units where $8\pi G=c=1$. Since $\sqrt{-g}\sim Na^3$ one can integrate (\ref{action}) over the spatial variables and substitute $t\to i\tau$, thus reducing (\ref{action}) to Euclidean action:
\begin{equation} \label{Eaction}
I=\int N d\tau \left( -\frac{a{a'}^2}{N^2} +\frac{{w'}{a'}a^2}{2N^2}+\frac{w a^3\rho}{3} \right),
\end{equation}
where ${}'=d/d\tau$. Then in the gauge where $N=1$ we have Hamiltonian constraint
\begin{equation}
H=\frac{\delta I}{\delta N}-(w+1)\frac{a^3\rho}{3}=0, \label{H5}
\end{equation}
and the momenta conjugate to $a$ and $w$:
\begin{equation}
\pi_a=i \left(\frac{1}{2}\dot \omega a^2 - 2 a \dot a\right), \qquad \pi_w=i\frac{1}{2}\dot a a^2.
\end{equation}
However, the rules of quantum description calls for them to be replaced by the corresponding quantum operators:
\begin{equation} \label{momenta}
\pi_a\to -i\frac{\partial}{\partial a},\qquad \pi_w\to -i\frac{\partial}{\partial w}.
\end{equation}
By making this change inside of the Hamiltonian constraint we will finally obtain the Wheeler-DeWitt equation:
\begin{equation} \label{WDW}
\left(\frac{\partial^2}{\partial w^2
}-\frac{1}{2}a\frac{\partial^2}{\partial w\partial
a}+\frac{3}{4}a^6\rho\right)\Psi(w,a)=0,
\end{equation}

For all intents and purposes, (\ref{WDW}) is a hard nut to crack even for the simplest physically meaningful boundary conditions. However, as we shall see, this fortress has a hidden back door; in fact, in the next section we are going to show that by employing the isospectral symmetries like the Moutard and Darboux transformations, it is possible to construct the exact solutions of (\ref{WDW}). Or, to say it in a more rigorous way, there is a simple method of obtaining the ''potentials'' $\rho(w,a)$ for which the equation (\ref{WDW}) can be solved exactly.

%Then using the approach from the \cite{Gonzalez-3} we offer the generalized Wheeler-DeWitt equation for wave function of the universe (without condition ${\ddot w}=0$ and $k=0$) and show that universes in multiverse must be in the state with $w=-1/3$ just after their big trips.

\section{Moutard transformation for Wheeler-DeWitt equation}

Let us introduce two new independent variables $u$ and $v$:
\begin{equation}
u=\frac{w}{2}+\ln a,\qquad v=-\ln a, \label{uv}
\end{equation}
and
$$
w=2(u+v),\qquad a={\rm e}^{-v}.
$$
As $v\to+\infty$ we'll have $a\to 0$ and $v\to-\infty$ would imply $a\to+\infty$. The case
$$
u+v=-\frac{1}{2},
$$
corresponds to a dS universe and when
$$
u+v<-\frac{1}{2},
$$
we automatically have a phantom equation of state.

By direct computation
$$
\frac{\partial}{\partial w}=\frac{1}{2}\frac{\partial}{\partial
u},\qquad \frac{\partial}{\partial
w}-\frac{1}{2}a\frac{\partial}{\partial
a}=\frac{1}{2}\frac{\partial}{\partial v}.
$$
Therefore, in terms of the new variables the equation (\ref{WDW}) has the form
\begin{equation}
\left(\frac{\partial^2}{\partial u\partial
v}+U(u,v)\right)\Psi(u,v)=0, \label{Wheeler-DeWitt-1}
\end{equation}
where
$$
U(u,v)=3 {\rm e}^{-6v} \rho(u,v).
$$
The equation (\ref{Wheeler-DeWitt-1}) admits the Moutard transformation (MT). Namely, let $\Psi$ and $\Phi$ be two solutions of
the (\ref{Wheeler-DeWitt-1}):
\begin{equation}
-\frac{1}{\Psi}\frac{\partial^2\Psi}{\partial u\partial
v}=-\frac{1}{\Phi}\frac{\partial^2\Phi}{\partial u\partial
v}=U(u,v). \label{solutions}
\end{equation}
Define a 1-form $d\theta[\Psi;\Phi]$ such that
\begin{equation}
d\theta[\Psi;\Phi]= du\left(\frac{\partial\Psi}{\partial u
}\Phi-\frac{\partial\Phi}{\partial u
}\Psi\right)-dv\left(\frac{\partial\Psi}{\partial v
}\Phi-\frac{\partial\Phi}{\partial v }\Psi\right), \label{form}
\end{equation}
and
\begin{equation}
\theta[\Psi;\Phi]=\int_{\Gamma}d\theta[\Psi;\Phi]. \label{theta}
\end{equation}
Note that since by definition both $\Psi$ and $\Phi$ are solutions of (\ref{solutions}), the one-form is closed, i.e.
$$
\frac{\partial^2\theta[\Psi;\Phi]}{\partial u\partial
v}=\frac{\partial^2\theta[\Psi;\Phi]}{\partial v\partial u},
$$
and thus the shape of the contour of integration $\Gamma$ is irrelevant.

The MT has the form
\begin{equation}
\Psi\to\Psi^{(1)}=\frac{\theta[\Psi;\Phi]}{\Phi}, \label{Psi1}
\end{equation}
\begin{equation}
U\to U^{(1)}=U+2\frac{\partial^2}{\partial u\partial v}\ln\Phi.
 \label{U1}
\end{equation}
It means that
\begin{equation}
-\frac{1}{\Psi^{(1)}}\frac{\partial^2\Psi^{(1)}}{\partial u\partial
v}=U^{(1)}(u,v). \label{dressed}
\end{equation}
It is worth pointing out at this step that
\begin{equation}
\Phi^{(1)}=\frac{1}{\Phi}, \label{Phi1}
\end{equation}
rather then zero.

The Moutard transformation (\ref{Psi1}), (\ref{Phi1}) , (\ref{U1})
can be iterated several times, and the result can be expressed via
corresponding Pfaffian forms.
\newline
{\bf Example 1.}

Let $U=0$, therefore
$$
\Psi=A(u)+B(v),\qquad \Phi=\alpha(u)+\beta(v),
$$
with arbitrary functions $A(u)$, $B(v)$,  $\alpha(u)$ and
$\beta(v)$. Using (\ref{U1}) we get
\begin{equation}
U^{(1)}=-\frac{2\alpha'(u)\beta'(v)}{(\alpha(u)+\beta(v))^2}.
\label{dress-U1}
\end{equation}
Using (\ref{Psi1}) we can calculate the solution of equation (\ref{dressed}) with the potential (\ref{dress-U1}), that would contain two arbitrary functions $A(u)$ and $B(v)$:
$$
\Psi^{(1)}=\frac{2(\beta(v)A(u)-B(v)\alpha(u))+\theta_1(u)+\theta_2(v)}{\alpha(u)+\beta(v)},
$$
where
$$
\theta_1(u)=\int du\left(A'(u)\alpha(u)-A(u)\alpha'(u)\right),
$$
$$
\theta_2(v)=\int dv\left(B(v)\beta' (v)-B'(v)\beta (v)\right).
$$
The second solution (without arbitrary functions) is given by the formula (\ref{Phi1}):
\begin{equation}
\Phi^{(1)}=\frac{1}{\alpha(u)+\beta(v)}. \label{uh}
\end{equation}
One simple yet interesting example follows if one takes
\begin{equation}
\alpha(u)=\cosh(u),\qquad \beta(v)=\cosh(v). \label{case}
\end{equation}
In this case the expression (\ref{uh}) describes the normalizable wave function:
$$
\int_{-\infty}^{+\infty}\int_{-\infty}^{+\infty}dudv|\Phi^{(1)}|^2<\infty,
$$
unlike the original solution $\Phi$, which does not belong to $L^2(u,v)$ for any (non-zero) functions $\alpha(u)$ and $\beta(v)$. Furthermore, the new solution clearly has a maximum at $u=v=0$, i.e. at $a=1$ and $w=0$ (which also happens to be a saddle point of potential $U^{(1)}$). Thus, if we operate in framework of this model we would have to find ourselves with large probability in a dust universe with a finite value of scale factor.

{\bf Example 2. Dressing of the dS universe}

Let $\rho=\Lambda/3={\rm const}>0$, so $U(u,v)={\rm e}^{-6v}\Lambda$. The solution of the (\ref{Wheeler-DeWitt-1}) has a general form
$$
\displaystyle{
\Psi(u,v)=\int d\kappa f(\kappa){\rm e}^{\kappa u+\frac{\Lambda}{6\kappa}{\rm e}^{-6v}},}
$$
with an arbitrary integrable function $f(\kappa)$. Let us consider a special case:
\begin{equation}
\displaystyle{
\Psi(u,v)=\sum_{j=1}^n c(\kappa_j) {\rm e}^{\kappa_j u+\frac{\Lambda}{6\kappa_j}{\rm e}^{-6v}},}
\label{reshenie}
\end{equation}
with real-valued $\kappa_j$. For the sake of simplicity, let $n=2$, $c(\kappa_j)=1/2$, $\kappa_2=-\kappa_1=-\kappa$. Then
\begin{equation}
\label{psps}
\Psi(u,v)=\cosh(\gamma), \qquad \gamma(u,v)=\kappa u+\frac{\Lambda}{6\kappa}{\rm e}^{-6v},
\end{equation}
and we can write the result of the Moutard dressing as
\begin{equation}
\Psi^{(1)}=\frac{1}{\cosh(\gamma)},
\label{norm}
\end{equation}
and the density $\rho(u,v)$ as
\begin{equation}
\rho^{(1)}=\Lambda\left(1-\frac{2}{\cosh^2\gamma}\right).
\label{r1}
\end{equation}
Note, that in this case the maximal probability distribution will not be localized at one point in the $(a, w)$-phase space, but instead we have a continuous curve $a(w)$ that solves the equation $\gamma(u,v)=0$.

These examples show that, unless we have a very special case (like $U=0$), the eigenfunctions of the Wheeler-DeWitt equation (\ref{WDW}) would in general have not just one but many (possibly infinitely many) maximums in the $(a, w)$ space, thus leaving us with the burden of choosing the appropriate $w$ that the universe should attain after the big trip. Of course, there exist different possibilities for such a choice -- for example, using the final anthropic principle or restricting ourselves to a discrete set of possible values of $\omega$ (that is treated as a quantum-mechanical parameter); nevertheless, it is also possible that such a freedom might actually imply the possible deficiency of the equation itself. In fact, as we shall see below, once we take into account possibility of $\ddot{w}\ne 0$ and $k \ne 0$, the resulting equation would appear to favor one special value of $w$, namely: $w=-1/3$.

\section{Generalized Wheeler-DeWitt equation and the big trip}

The standard definition of a big trip implies that it is a cosmological event thought to occur in the future during which the entire universe is engulfed inside of a gigantic wormhole (hence, the term ``big''!), traveling through it along space and time.

However, it would appear that yet another ``Big'' cosmological event might be possible; an event that, by analogy, we would like to call a ``Big Meeting''.

In order to get to it, we will use the approach from \cite{Gonzalez-3} with alteration: we would not impose the restrictions ${\ddot w}=0$ and $k=0$. As before, we are assuming that $\rho=\rho(w,a)$. Differentiating the Einstein-Friedmann equation $H^2=\rho/3-k/a^{2}$ with respect to $t$ will get us a different expression for the scalar curvature ${\hat R}$:
\begin{equation}
{\hat R}=R+\frac{1}{{\dot a}a^2}\left(9(1+w)({\dot
a}^3+k\dot{a})+a^3{\dot\rho}\right), \label{hatR2}
\end{equation}
where
$$
R=\frac{6({\dot a}^2+a{\ddot a}+k)}{a^2},\quad {\dot\rho}=\frac{\partial\rho}{\partial a}{\dot
a}+\frac{\partial\rho}{\partial w}{\dot w}.
$$
The corresponding euclidian action is
\begin{equation}
I=\int d\tau
N\left[\frac{6aa'^3-aF(a,a',w,w')}{2a'N^2}+3ka+a^3w\rho\right],
\label{Eaction2}
\end{equation}
where
$$
\begin{array}{ll}
F(a,a',w,w')=9(1+w)(a'^3-ka')-\\
-a^3\left(\frac{\partial\rho}{\partial
a}a'+\frac{\partial\rho}{\partial w}w'\right).
\end{array}
$$
Introducing the momenta according to (\ref{momenta}) and redefining the (classical) Hamiltonian via $\pi_a$, $\pi_w$, $a$ and $w$ will provide us with a generalized Wheeler-DeWitt equation:
\begin{equation}\begin{array}{ll}
4\left(a\frac{\partial\rho}{\partial
a}+2\rho+3(1+3w)\frac{k}{a^2}\right)\frac{\partial^2\Psi(a,w)}{\partial
w^2}-\\
-4a\frac{\partial\rho}{\partial
w}\frac{\partial^2\Psi(a,w)}{\partial a\partial w}=-9\left(\frac{\partial\rho}{\partial w}\right)^2a^6(1+3w)\Psi(a,w).
\end{array}
\label{WD6}
\end{equation}
The WDE (\ref{WD6}) is different from the WDE (\ref{WDW}), obtained in
\cite{Gonzalez-3} since the absence of condition ${\ddot w}=0$ has made a further simplification of the actions (\ref{action}), (\ref{Eaction}) not possible. As we shall see, it is this new equation (\ref{WD6}) that results in a new effect -- the ``Big Meeting''.

First at all, lets consider the special case $w=-1/3$. The right hand side of the (\ref{WD6}) will be equal to zero. Moreover, we will have $\rho=\rho(a,w=-1/3)\sim a^{-2}$, and thus
$$
a\frac{\partial\rho}{\partial a}+2\rho=0,
$$
which reduces the WDE (\ref{WD6}) to
\begin{equation}
\frac{\log a}{a}\frac{\partial^2\Psi}{\partial a\partial w}=0.
\label{uh7}
\end{equation}
Using the power series expansion of function $\Psi(\cdot,\omega)$ around the point $(\cdot, -1/3)$
$$
\Psi(a,w)=\Psi(a,-1/3)+\sum_{n=1}^{\infty}\frac{1}{n!}c_n(a)\left(w+\frac{1}{3}\right)^n,
$$
and substituting it into (\ref{uh7}) would result in $dc_1(a)/da=0$, and thus in
\begin{equation}
\frac{\partial \Psi(a,w)}{\partial w}|_{w=-1/3}=c_1={\rm const}.
\label{8}
\end{equation}
On the other hand, equation (\ref{uh7}) is invariant with respect to the transformation
\begin{equation}
\Psi(a,w)\to\Psi(a,w)-f_1(a)-f_2(w), \label{9}
\end{equation}
for arbitrary (differentiable) functions $f_{1,2}$. Substituting (\ref{9}) into the (\ref{8}) and choosing $df_2(w)/dw=c_1$ at $w=-1/3$ we get without
loss of generality
\begin{equation}
\frac{\partial \Psi(a,w)}{\partial w}|_{w=-1/3}=0. \label{extr10}
\end{equation}
Therefore, in the case of general position, for any {\bf fixed} $a$ the function $\Psi(a,w)=\Phi_a(w)$ has an extremum at $w=-1/3$. Moreover, the condition that function $\Psi$ has to be normalizable implies that the probability distribution $|\Psi(a,w)|^2$ for any given value of the scale factor has a peak at $w=-1/3$. In order to prove this lets return to equation (\ref{WD6}). Consider its solutions that satisfy the property
\begin{equation}
\frac{\partial \Psi(a,w)}{\partial w}|_{w=w_0}=0, \label{extr11}
\end{equation}
for any given $a$ and some fixed $w_0={\rm const}$. Then $\rho(a,w_0)=\rho_0=C^2a^{-3(w_0+1)}$ ($C={\rm const}$) and at point $(\cdot, w_0)$:
\begin{equation}
a\frac{\partial\rho_0}{\partial a}|_{w=w_0}=-3(w_0+1)\rho_0, \label{12}
\end{equation}
\begin{equation}
\frac{\partial\rho_0}{\partial w_0}|_{w=w_0}=-3\log a\rho_0. \label{13}
\end{equation}
Besides
\begin{equation}
\frac{\partial^2\Psi}{\partial a\partial
w}|_{w=w_0}=\frac{\partial}{\partial a}\left(\frac{\partial
\Psi(a,w)}{\partial w}|_{w=w_0}\right)=0. \label{14}
\end{equation}
Substituting (\ref{12}), (\ref{13}) and (\ref{14}) into the (\ref{WD6}) we get
\begin{equation}\begin{array}{ll}
(1+3w_0)(4a^{-3(w_0+1)}(1-3kC^{-2}a^{1+3w_{0}})\frac{\partial^2\Psi}{\partial
w^2}|_{w=w_0}\\
-81C^2a^{-6w_0}\log^2a \Psi)=0. \label{15}
\end{array}
\end{equation}
Using (\ref{15}) one can conclude that either $w_0=-1/3$ or we have to satisfy the following equation
\begin{equation}\begin{array}{ll}
(1-3kC^{-2}a^{1+3w_{0}})\frac{\partial^2\Psi}{\partial
w^2}|_{w=w_0}=\\
=\frac{81 C^2}{4}a^{-3(w_0-1)}\log^2a \Psi. \label{16}
\end{array}
\end{equation}
If $k=0$ the general solution of the (\ref{16}) has the form
\begin{equation}
\Psi(w_0,a)=c_1 I_0(z)+c_2 K_0(z), \label{solution17}
\end{equation}
where $c_{1,2}$ are arbitrary constants, $z=3Ca^{3(1-w_0)/2}$, $I_0$ and $K_0$ are the Bessel functions of the first and second kind correspondingly. This function has to be normalizable:
\begin{equation}
\int_0^{\infty} da|\Psi(w_0,a)|^2<+\infty,
\label{integr}
\end{equation}
which implies that $c_1=0$. Substituting (\ref{solution17}) into the (\ref{14}) results in
$$\begin{array}{ll}
\frac{\partial\Psi}{\partial w}|_{w=w_0}=\\
=-\frac{9 C}{2}a^{3(1-w_0)/2}\log
a K_1\left(3 C a^{3(1-w_0)/2}\right)=0,
\end{array}
$$
which can only be satisfied for arbitrary $a$ if and only if $w_0=1$ and $C$ is the solution of the equation
$$
CK_1(3C)=0.
$$
But if $w_0=1$ then the wave function $\Psi={\rm const}$ (and the same will be the case for the density $\rho$ (see (\ref{WD6}))).
But the only normalizable function of this type is $\Psi=0$. Therefore, the only possible non-trivial solution that comply with (\ref{extr11}) in framework of normalizable wave functions of the universe correspond to the case $w_0=-1/3$.

{\em The end of the proof}.

Now lets say a few words on account of possible existence of other peaks $M_*=(a_*,w_*)$ on the graph of solutions of the (\ref{WD6}) that satisfy
$$
\frac{\partial\Psi}{\partial
a}|_{M_*}=\frac{\partial\Psi}{\partial w}|_{M_*}=0,
$$
and
\begin{equation}
\Delta=\left[\frac{\partial^2\Psi}{\partial
w^2}\frac{\partial^2\Psi}{\partial
a^2}-\left(\frac{\partial^2\Psi}{\partial w\partial
a}\right)^2\right]_{M_*}>0. \label{det18}
\end{equation}
Surprisingly, it is possible to eradicate these extrema. To show it, one should take into account the dimensionless variables in expressions
like $\log a$. Keeping in mind that our universe was born via big trip from the paternal universe with scale factor $L$, we have to replace $\log a\to\log(a/L)$. On the other hand, we expect that right after the big trip the initial value of the scale factor $a_i$ will be $a_i\sim
L$. Lets consider such a universe with $w={\rm const}$. In order to estimate the probability to find this universe just after the big trip (that is with $a_i\sim L\sim a_*$) one must use the WDE (\ref{WD6}) which reduces to a simple form
$$
-4a^{-3(w+1)}(3w+1) \frac{\partial^2\Psi}{\partial w^2}|_{M_*}=0.
$$
If $w\ne -1/3$ then (see (\ref{det18}))
$$
\Delta=-\left(\frac{\partial^2\Psi}{\partial w\partial
a}\right)^2 |_{M_*}<0,
$$
and we have no extremum at all.

Note one interesting feature of a universe with $w=-1/3$. If $w=-1/3$ the solution for scale factor has a simple form
$$
a(t)=At+a_{0},
$$
where $A$ and $a_{0}$ are constants. For the expanding universe $A>0$. It is possible to join the metric with this scale factor to any FRW universe at any time in the expanding phase. Consider for example a flat de Sitter universe:
$$
a(t)=C\exp(\lambda t)
$$
where $C=const$ and $\lambda=\sqrt{8\pi G \rho_{\Lambda}/3}$.  The matching conditions for arbitrary $t=\tau$ require continuity of the scale factor together with its first derivatives:

$$
a(\tau)=C\exp(\lambda \tau)=A\tau+a_{0},
$$

$$
\dot{a}(\tau)=C\lambda\exp(\lambda\tau)=A.
$$
For $a_{0}$ we have $a_{0}=C(1-\lambda\tau)\exp(\lambda\tau)$. Therefore the joining is trivial: the continuity condition of continuity of $\dot{a}(t)$ uniquely determines the parameter $A$ and the continuity condition of $a(t)$ does the same for the constant $a_{0}$.

\section{Conclusion}

The simple method based on Moutard transformation for constructing exact solutions of the Wheeler-DeWitt equation (\ref{WDW}) is presented. A direct application of this method has demonstrated that in general, unless some kind of additional restriction on the possible values of $w$ is imposed, we would have to face a problem of somehow choosing a suitable $w$ from (in general) infinite set of possibilities whenever we try to describe the universe after the big trip.

However, it appears that should one consider the generalized form of Wheeler-DeWitt equation (\ref{WD6}) instead, the conclusion would be drastically different. In this case the probability distribution has the peak(s) at $w=-1/3$ for any $a$ provided that we consider the universe right after big trip. By analogy with the ``big trip'' this result can be called a ``Big Meeting''. What it means is that in multiverse the vast majority of universes experiencing the big trip should end up in a state with $w=-1/3$ right afterwards.

\end{document}